\documentstyle[preprint,aps,prl,epsf]{revtex}   
\begin{document}
\tightenlines 
\title{{\bf Stretching of proteins in a force-clamp}}
\author{{\bf Piotr Szymczak $^1$ and Marek Cieplak$^2$}}

\address{
$^1$Institute of Theoretical Physics, Warsaw University,
ul. Ho\.za 69, 00-681 Warsaw, Poland\\
$^2$Institute of Physics, Polish Academy of Sciences,
Al. Lotnik\'ow 32/46, 02-668 Warsaw, Poland}

\maketitle



\vspace*{1cm}

\begin{abstract}
{Mechanically induced protein unfolding in the force-clamp
apparatus is shown, in a coarse-grained model of ubiquitin,
to have lognormal
statistics above a treshold force and exponential below it.
Correspondingly, the mean unfolding time is slowly
varying and exponentially decreases as a function of the force.
The time dependencies
of the end-to-end distances are also distinct.
The time sequence of unfolding events weakly depends on force and much of it resembles that
for stretching at
constant speed. A more complicated
time dependence arises for integrin.
}
\end{abstract}

\newpage

The atomic force microscopy (AFM) provides a convenient tool to to mechanically probe
biomolecules.
Usually, it is employed for
stretching at a constant speed.
The properties of the force of resistance to
stretching, as measured as a function of the AFM tip displacement,
yield information about
the elastic structure of the biomolecule.
Examples of such studies involve titin \cite{Gaub,Marszalek,Fowler2002}
and DNA \cite{Heslot}.
Recently, a new variant of AFM has been developed: a force-clamp.
It allows to maintain a
constant pulling force on the protein while monitoring the end-to-end distance, $L$, as a
function of time, $t$.
This technique has been used to study unfolding of
the domains of titin \cite{clampober} and polyubiquitin \cite{FernandezLi,Schlierf}.
In the latter case, the end-to-end length of the protein subject to a
constant force was found to increase in a step-wise
manner with each step corresponding to unwinding of a single ubiquitin chain in
agreement with a simple two-state, all-or-none
model of unfolding. The ensemble averaged length of a protein is well
described by a single exponential with a characteristic time
$t_{unf}$ \cite{Schlierf}. The logarithm of $t_{unf}$ was
shown to depend linearly on the force $F$.
In the present study we investigate the generality of these results.
In particular, we show that the average unfolding time needs not to be
exponential (especially at large forces) and we provide example
of a protein with more than two steps of unfolding at low forces.\\

One of the results that comes from the constant speed experiments is an
estimate of the maximum force, $F_{max}$, that is needed to unravel
the protein. This information can be obtained even from a single pulling
trajectory. In a force-clamp experiment, however, a single trajectory,
as represented by the $L(t)$ trace, is not very revealing because
the sudden jumps in $L$ take place at seemingly random moments.
Furthermore, the appearance of the trace
does not tell whether it has been measured above or below
$F_{max}$, especially if $F$ is close to $F_{max}$.
A physical information can be gleaned only by considering ensembles
made of many trajectories and determining properties of distributions
of the unfolding times.\\

Here, we provide a theoretical assessment of the constant-force
unfolding experiments from the perspective of a statistical analysis
of many trajectories. We focus on ubiquitin and
derive distributions of the unfolding times as a function
of the force and show that their nature switches from
exponential to lognormal on crossing the treshold value $F_{max}$.
Furthermore, we demonstrate that for $F > F_{max}$
the mean unfolding time, $t_{unf}$, is almost independent of the force, but it
grows exponentially on lowering the force for $F < F_{max}$. The fact
that on crossing $F_{max}$ the kinetics of unfolding is changed
is not surprising in itself. However, the purpose of this
paper is to quantify the character of the
statistical change. In particular, we
demonstrate existence of different time dependencies of the ensemble
averaged $L$ for above and below $F_{max}$.\\

Proteins, when stretched at a constant speed, usually exhibit several
force maxima and one of them is often clearly dominant. In the case of
ubiquitin and the I27 domain of titin the dominant peak is the
first significant maximum that comes as a function of the
displacement, $d$ \cite{thermtit,cieplakmarszalek}.
However, one can find proteins, like integrin, for which the dominant
peak force corresponds to a later maximum. We show that this
feature complicates the behavior of the unfolding characteristics
still further.\\

It is difficult to generate a
theoretical account of unfolding in the force-clamp apparatus
using all-atom simulations. Not only the time scales required for
single trajectory unfolding simulations
are orders of magnitude too long but also there is
an intrinsic necessity to consider many unfolding processes
in order to discuss ensembles. Go-like models \cite{Goabe,Stakada}
offer a rescue in this context. We focus on ubiquitin and integrin as
case studies and model the two proteins in the Go-like fashion.
We follow the implementation as
outlined in refs. \cite{biophysical,thermtit} and, specifically
for ubiquitin, in ref. \cite{cieplakmarszalek}. The latter paper
discusses stretching at constant speed. The model consists of a
chain of self-interacting C$^{\alpha}$ atoms that are
tethered by harmonic potentials with minima at 3.8 {\AA}.
The other interactions are
selected so that the global energy minimum agrees with
the experimentally established native conformation.
The interactions, called contacts, can be divided into native and
non-native kinds by checking atomic overlaps
in the native conformation as described
by Tsai et al. \cite{Tsai}. In order to prevent emergence of
entanglements, the non-native contacts are endowed with
a hard core repulsion at a distance of $\sigma$=$5$ {\AA}.
The native contacts are described by the
Lennard-Jones potentials
$V_{ij} =
4\epsilon \left[ \left( \frac{\sigma_{ij}}{r_{ij}}
\right)^{12}-\left(\frac{\sigma_{ij}}{r_{ij}}\right)^6\right]$, where
the length parameters $\sigma _{ij}$ are chosen so that the potential
minima correspond to the native distances between
the C$^{\alpha}$ atoms $i$ and $j$.
The energy parameter, $\epsilon$, is taken
to be uniform and its effective value for titin and ubiquitin
appears to be of order 900 K so the reduced temperature,
$\tilde{T}=k_BT/\epsilon \;$ of 0.3
($k_B$ is the Boltzmann constant and $T$ is temperature)
should be close to the room temperature value \cite{cieplakmarszalek}.
Unless mentioned otherwise, all results are obtained for this
value of $\tilde{T}$.
In our stretching simulations, the N-terminus of the protein is
attached to harmonic springs of elastic
constant $k$=0.06 $\epsilon /${\AA}$^2$.
The C-terminus is pulled by a force, $F$. In the constant speed
simulations, the C-terminus is attached to another harmonic spring
with the same $k$ as at the N-terminus.
The other end of the C-terminus spring moves
at a speed $v_p$=0.005 $\AA /\tau$, where
$\tau =\sqrt{m \sigma^2 / \epsilon}$ is the characteristic time unit
and $m$ is the average mass of the amino acids. Structural frustration is reduced by
introducing a fourbody
chirality term in the Hamiltonian~\cite{Kwie1}\\

Thermostating is provided by the Langevin noise which also
mimics random kicks by molecules of the implicit solvent.
An equation of
motion for each C$^{\alpha}$ reads
$m\ddot{{\bf r}} = -\gamma \dot{{\bf r}} + F_c + \Gamma $, where
$F_c$ is the net force on an atom due to the molecular potentials
and $\Gamma$ is a Gaussian noise term with dispersion
$\sqrt{2\gamma k_B T}$.
The damping constant $\gamma$ is taken to be equal to $2m/\tau$
and the dispersion of the random forces is equal to
$\sqrt{2\gamma k_B T}$.
This choice of $\gamma$ corresponds to a situation in which the
inertial effects are negligible \cite{biophysical}. In order to
make the damping as effective as in water, $\gamma$ should be about
25 times larger and the resulting time scales would become
about 25 times longer \cite{biophysical}.
The equations
of motion are solved by a fifth order predictor-corrector scheme.\\

Three panels of Figure 1 show examples of the $L(t)$ trajectories for
ubiquitin, a ubiquitin dimer, and integrin. The time evolution consists of
steps in $L$ until the ultimate extension is reached. For ubiquitin,
there is just one step. For two-ubiquitin, there is a serial unwinding
of the domains and thus two steps in $L$. The example traces for ubiquitin
and two-ubiquitin are for the dimensionless force,
$\tilde{F}=F\sigma/\epsilon$,
of 2, which is below $\tilde{F}_{max}$ equal to $\sim$ 2.45.
In the case of integrin, the corresponding value of
$F_{max}$ is equal to $\sim 3.3$. The trajectories corresponding to the lower forces are not
developed to the full
extension within the scale of the figure. It is seen that single integrin
allows for multiple steps whereas single ubiquitin does not.
This difference is explained in the bottom right panel of Figure 1
which shows the constant speed results for the force determined against
displacement. The biggest force needed to unravel ubiquitin arises at the
beginning of the process. On the other hand, in the case of integrin
other modules unravel before the biggest force expense takes place.
Unwinding of these weaker modules gives rise to multiple steps
under the constant-force conditions.\\

Figure 2 shows what happens to $L$ when this quantity is averaged over
many trajectories. Instead of $L$ itself, we
consider a normalized, and dimensionless, $L'$, defined as
$\frac {L-L_f}{L_u-L_f}$, where $L_f$ and $L_u$ stand for the folded
and unfolded end-to-end distances respectively.
For $F > F_{max}$, the time dependence of $<L'>$
for ubiquitin is sigmoidal, with a point
of inflection (the top panel). On the other hand, for $F < F_{max}$, the
dependence is exponential with a single time scale given by $t_{unf}$
(the middle panel).
This panel also demonstrates that even though individual trajectories are
step-wise, an averaging over many trajectories leads to a quantity
which is governed by a single exponential, consistent with a two-state behavior
\cite{Schlierf}.\\

In the case of integrin (the bottom panel of Figure 2), we were not able to run
processes to achieve the full unwinding. However, the data averaged over
100 trajectories indicate disappearance of discrete steps
also in this case. We expect that the average behavior could
be fitted to a sum of two or three exponentials, depending
on the field, but demonstrating this would require significantly larger
statistics than available through our simulations.\\

The qualitative difference between the two force regions
becomes more transparent when one considers the average
unfolding times (Figure 3) and the distributions of the unfolding times
(Figure 4). Figure 3 shows that the average unfolding time depends
on the value of the force in the large force region very weakly,
possibly becoming force-independent asymptotically. However, it
displays an exponential dependence in the small force region.
The crossover is around $\tilde{F}$=2.3 which is close to $F_{max}$
as derived from one trajectory of the constant speed unwinding.
The value of $F_{max}$ decreases with the temperature which results
in shifting the region of the exponential dependence to smaller
forces as shown in the inset of Figure 3 for $\tilde{T}=0.6$
for which case $\tilde{F}_{max}$=0.88.\\

Figure 4 demonstrates that the large and small force regions
give rise to very different kinds of statistics of the individual
unfolding times, $t_u$. The distribution is lognormal above
$F_{max}$ and exponential below. The crossover between these two
statistics can be conveniently illustrated by
analyzing the moments of the distributions. In the insets of Fig. 4 we plot
certain combinations of the moments of $t_u$, $C$ and $C'$, defined in the
caption, such that $C$
is equal to 1 for the exponentially distributed variable,
whereas $C'$ is equal to 1 for the lognormal one. The
graphs clearly show a statistical crossover occuring
at $\tilde{F} \approx 2.2-2.3$, which is consistent with the result of the average unfolding
time analysis.
(see the arrows in the
insets of Figure 4).\\

The occurrence of the crossover
can be understood by observing that the unfolding time is actually
the sum of two terms: $t_1$ - the waiting time
for the unwinding to begin (which is distributed exponentially)
and $t_2$ - the duration of the unwinding process itself, which is largely
independent of the force (corresponding to the step width in Fig 1.).
For small forces, the unfolding time is
dominated by the waiting time which can be huge.
In contrast, for $F > F_{max}$ the
unwinding begins essentially immediately and $t_{unf}$ probes the fine
details of the unfolding process. The crossover between the two regimes
takes place when $t_1$ becomes comparable with $t_2$, which
corresponds to the streching force a bit below
$F_{max}$, just as it is seen in the simulations.\\

Finally, we discuss the unfolding scenarios. These can be characterized
\cite{thermtit} by determining average times at which a contact is broken
for the last time and plotted vs. the contact order, i.e. the
sequential separation $|j-i|$ between the $C^{\alpha}$ atoms that can
form a native contact. A contact is said to be broken if a distance
of $1.5 \sigma _{ij}$ is crossed. Figure 5 shows that the order of
rupturing events depends on $F$ rather weakly despite a significant
variation in the absolute time scales. This is consistent
with the two-stage picture outlined above.
It should be noted, however, that that there is a slight
change in the pattern of points on crossing $F_{max}$ that is easiest
to notice in the time sequencing of the hairpin (the solid squares)
at $|j-i|$ around 15.
It is interesting to point out that the order of unfolding events at
constant velocity, see Figure 6, is nearly the same as at constant force
(at the same temperature of $\tilde{T}=0.3$).
The most visible difference is that
the rupture of the helix 41-49 (the open pentagons) comes later than
the rupture of the bonds between (12-17) and (23-34) (open triangles).\\

In conclusion, results of our Go-like model highlight
fundamental force-driven crossover in the statistics of unfolding events
in experiments that involve the force-clamp AFM.

We appreciate help of Joanna I. Su{\l}kowska in identifying
integrin as having a possibly different behavior than ubiquitin.
This work was funded
by the Ministry of Science in Poland (grant 2P03B 03225).


\newpage
\centerline{FIGURE CAPTIONS}

\begin{description}

\item[Fig. 1. ]
The top two panels show examples of the
$L$ vs. time trajectories in a constant force protocol. The left and right
hand panels are for ubiquitin dimer and ubiquitin respectively.
In both cases, $\tilde{F}$=2, i.e., the force is below $F_{max}$.
The bottom left panel shows three examples of trajectories for integrin
at various forces.
 The lower two trajectories lead to full unfolding at
a significantly longer time than the scale in the figure.
The Protein Data Bank\cite{PDB} structure codes for ubiquitin
and integrin used here are 1ubq and 1ido respectively.
Ubiquitin consists of 76 amino acids and integrin of 192.
The bottom right  panel shows the force versus displacement traces
for stretching performed at constant speed. The trace for ubiquitin
is shifted upward by 1.7 to avoid overlap with the trace for integrin.
For this trace, $\tilde{F}_{max}$ is near 2.45.

\item[Fig. 2.]
The average normalized end-to-end length, $L'=\frac{L-L_f}{L_u - L_f}$,
as a function of time.
The top two panels are for single molecules of ubiquitin where the average is
over 380 trajectories. In this case, $L_u$=288$\AA$ and $L_f$=37.1$\AA$.
The top panel corresponds to
$F > F_{max}$ and the middle panel to $F < F_{max}$.
In the middle panel, two examples of individual step-wise trajectories
are also shown. The exponential fit corresponds to $t_{unf}$=0.32$\times 10^5 \tau$.
The data for integrin are shown in the bottom panel.
For integrin, $L_f$=10.82$\AA$ and $L_u$=699$\AA$.

\item[Fig. 3. ]
The mean unfolding time, $t_{unf}$ for ubiquitin,
as a function of the dimensionless force.
$\tilde{F}_{max}$ is close to 2.4 at $\tilde{T}=0.3$.
The solid line indicates the slope valid for $F < F_{max}$ and the
dotted line for $F > F_{max}$. The two lines intersect near $F_{max}$.
The inset shows the exponential regime a seen at a higher temperature
of $\tilde{T}=0.6$.

\item[Fig. 4.]
Distributions of unfolding times for ubiquitin.
The top panel is for $\tilde{F}=2.8$
and it corresponds to 1000 processes. The dotted line
indicates a fit to a lognormal distribution
$P(t/\tau)=\frac{1}{\sqrt{2\pi}\sigma (t-t_0)}
\exp{-\frac{ln^2(\frac{t-t_0}{m})}{2\sigma ^2}}$, with  $t_0$,
$\sigma$, and $m$ equal to 280 $\tau$, 0.37, and 105$\tau$ respectively,
and $t$ is short for the unfolding time $t_u$ such that $t_{unf}=<t_u>$.
The inset shows $C' = (\frac {<t^2>}{<t>^2})^3 \frac {<t>^3}{<t^3>}$
as a function of $\tilde{F}$. The unit value of $C'$,
seen for $\tilde{F} > 2.25$ is indicative of the lognormal distribution.
The bottom panel is for $\tilde{T}$=1.95, i.e. for an $F$ below $F_{max}$.
The data are based on 380 processes.
The dotted line shows a fit to an exponential distribution
$P(t)=\frac{1}{t_{unf}} \exp {-t/t_{unf}}$ with $t_{unf}=0.32\times 10^5 \tau$.
The inset shows $C=\frac{<t>}{\sigma _t}$, where $\sigma _t$ denotes the
dispersion in the distribution of $t$, as a function of $\tilde{F}$.
A unit value of $C$ is indicative of the exponential nature of the
probability distribution.

\item[Fig. 5.]
The stretching scenarios at constant force for ubiquitin for forces indicated.
The symbols assigned to specific contacts are the same in all panels.
Open circles, open triangles, open squares, open pentagons,
solid triangles, and solid
squares correspond to contacts (36-44)--(65-72), (12-17)--(23-34),
[(1-7),(12-17)]--(65,72), (41-49)--(41-49), (17-27)--(51-59), (1-7)--(12-17)
respectively. The crosses denote all other contacts.
The segment (23-34) corresponds to a helix. The two $\beta$-strand
(1-7) and ((12-17) form a hairpin. The remaining $\beta$-strands
are (17-27), (41-49), and (51-59).

\item[Fig. 6.]
The stretching scenario at constant pulling speed for ubiquitin.
The symbols used are as in Figure 5.

\end{description}

\begin{figure}
\epsfxsize=7in
\centerline{\epsffile{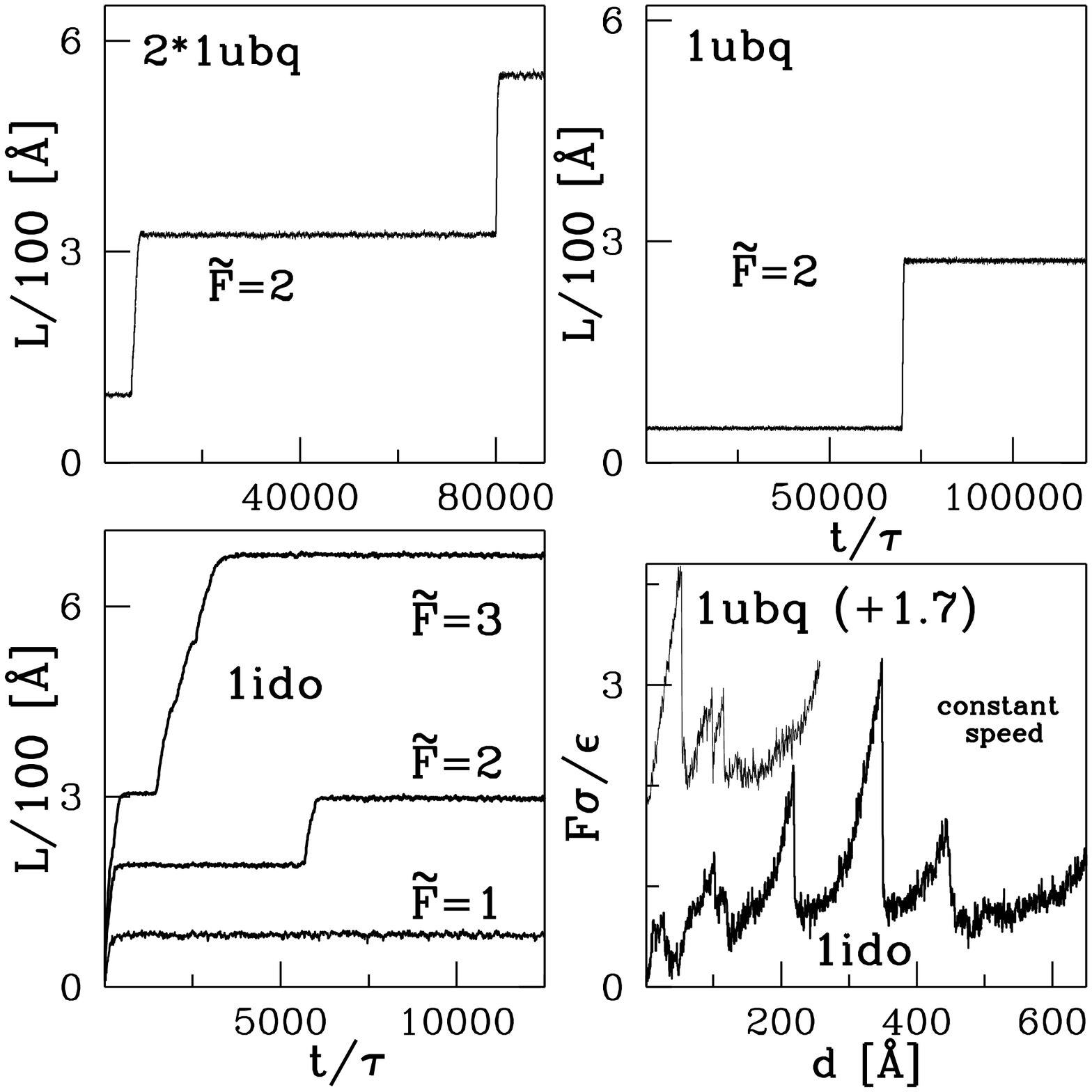}}
\vspace*{3cm}
\caption{ }
\end{figure}

\begin{figure}
\epsfxsize=7in
\centerline{\epsffile{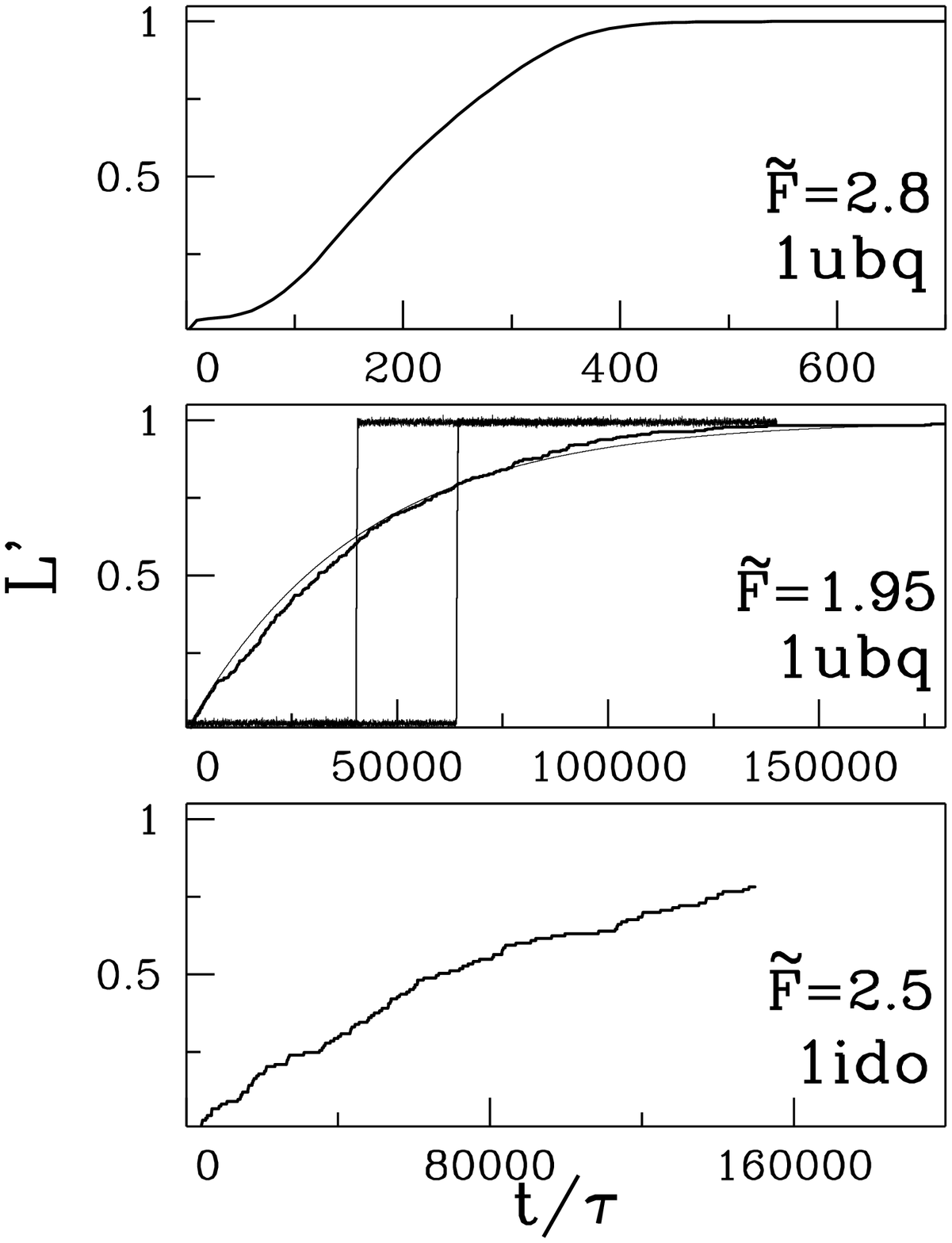}}
\vspace*{3cm}
\caption{ }
\end{figure}

\begin{figure}
\epsfxsize=7in
\centerline{\epsffile{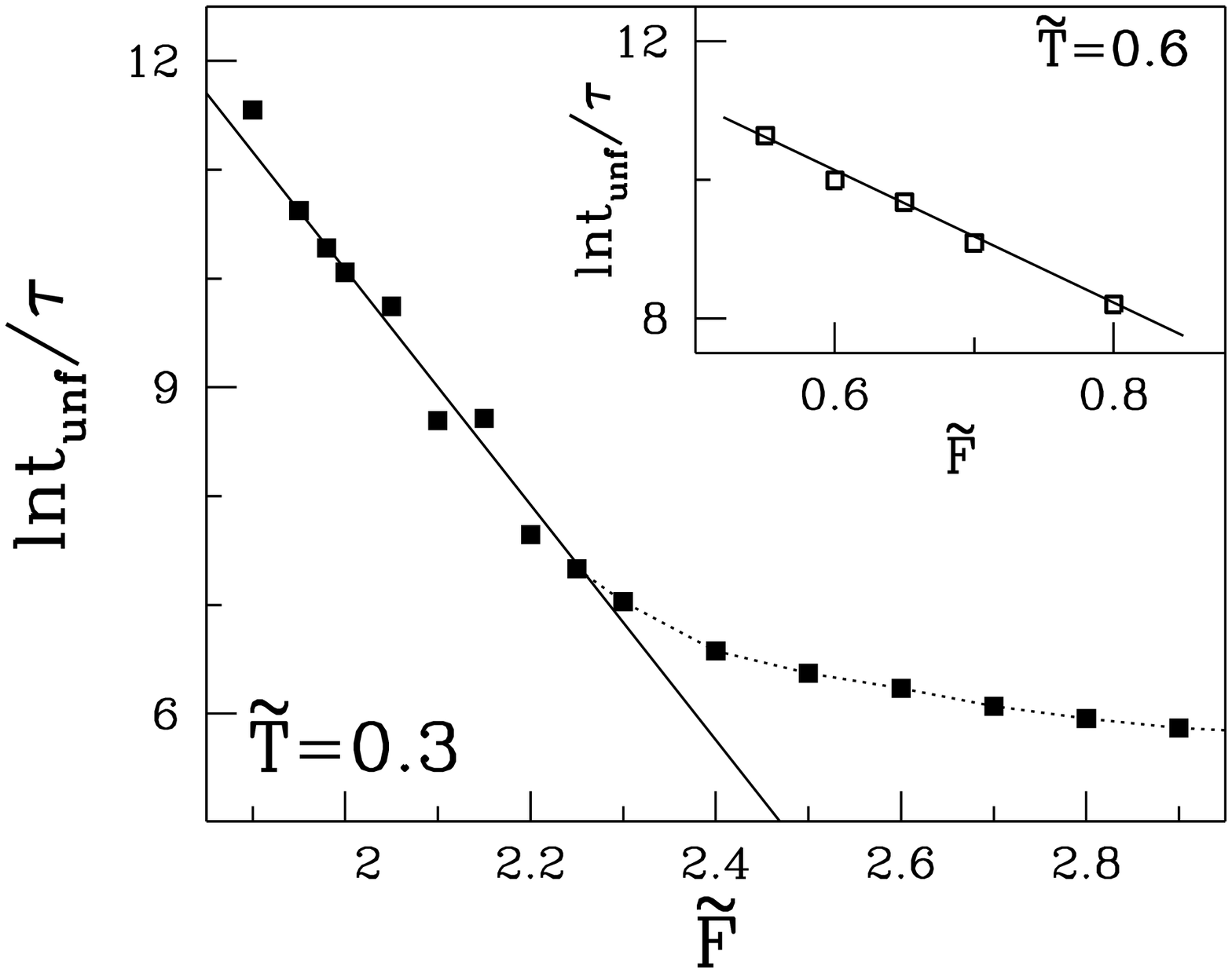}}
\vspace*{3cm}
\caption{ }
\end{figure}

\begin{figure}
\epsfxsize=6.6in
\centerline{\epsffile{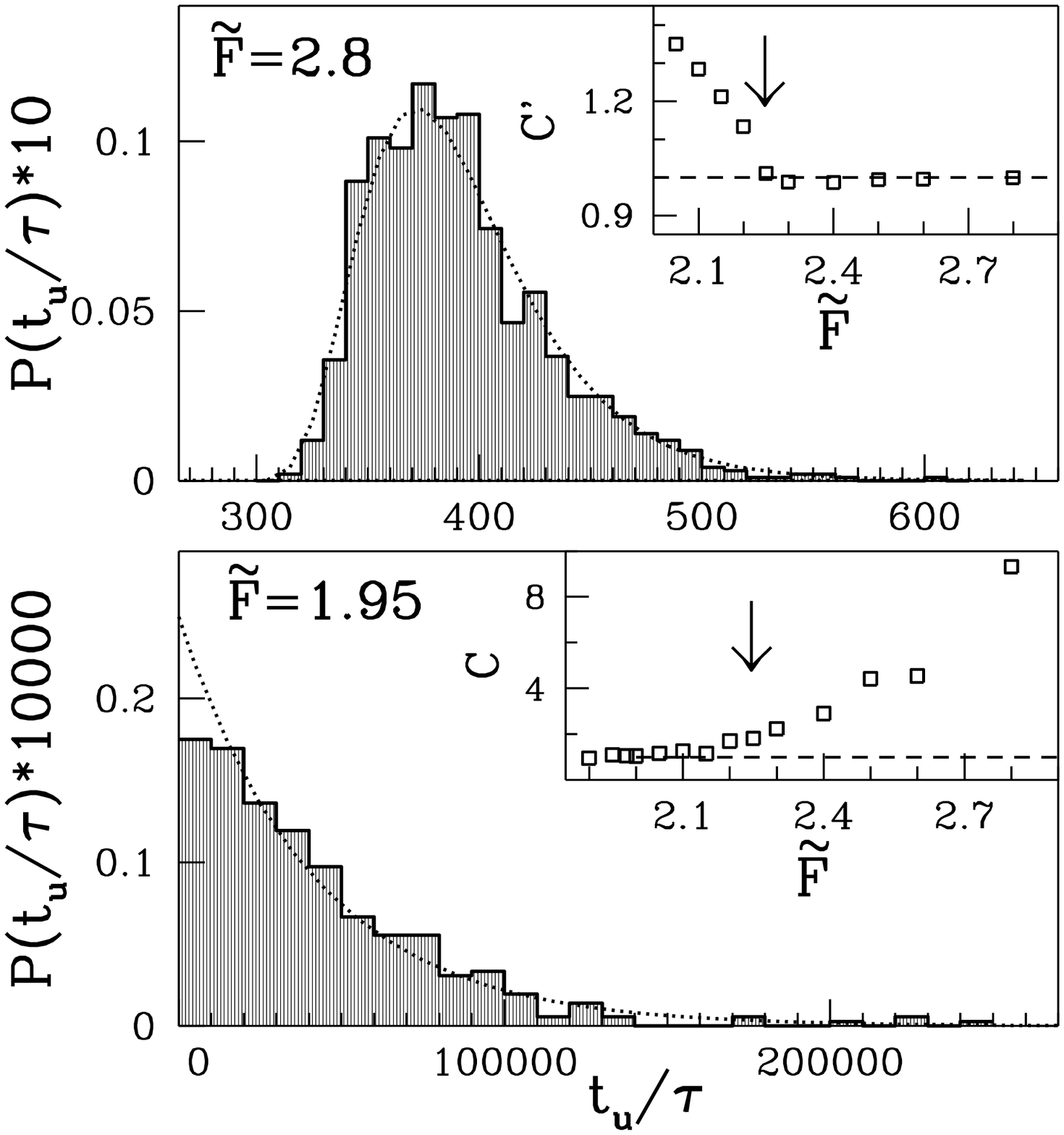}}
\caption{ }
\end{figure}

\begin{figure}
\epsfxsize=7in
\centerline{\epsffile{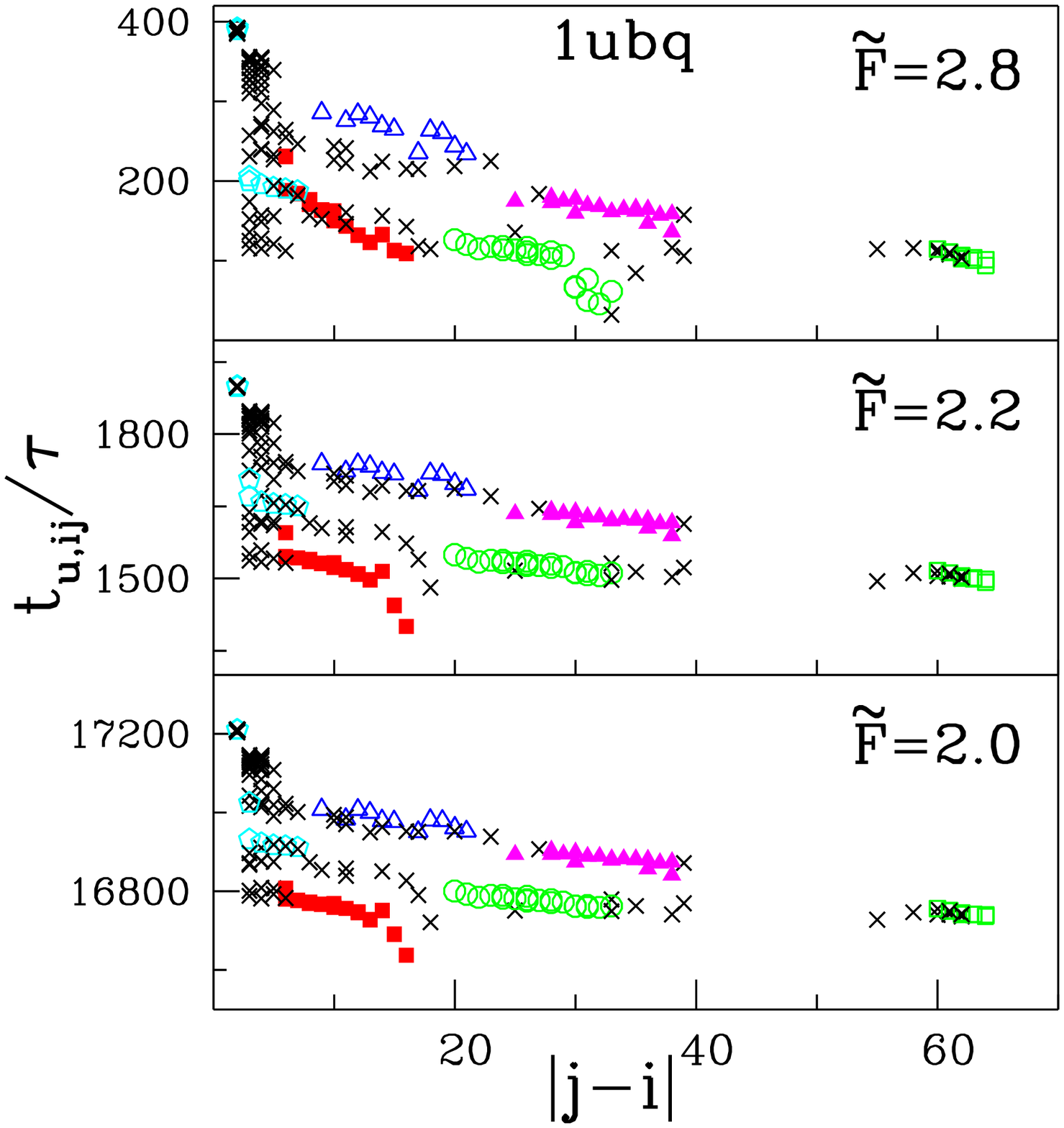}}
\vspace*{3cm}
\caption{ }
\end{figure}

\begin{figure}
\epsfxsize=6.6in
\centerline{\epsffile{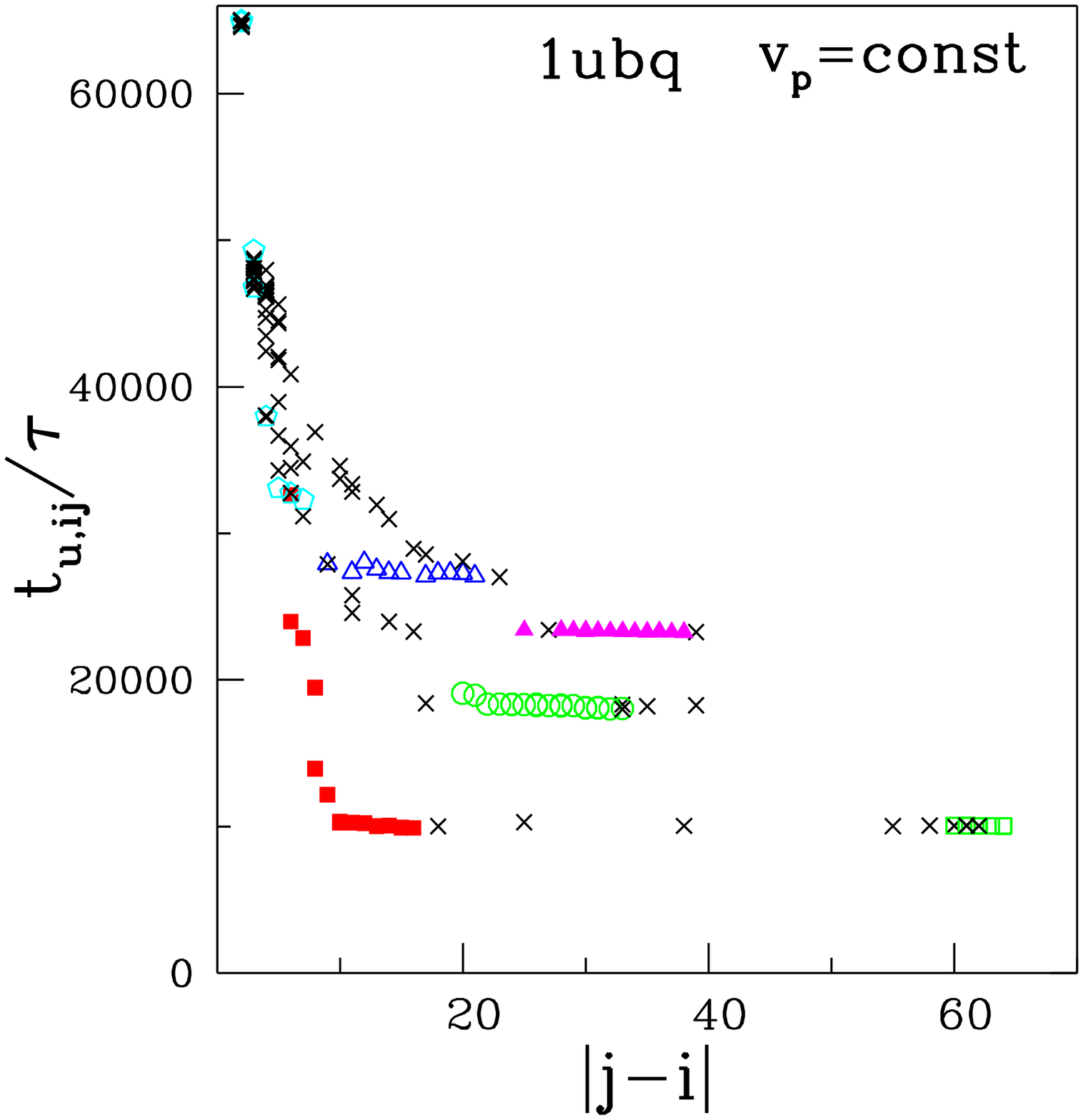}}
\caption{ }
\end{figure}

\end{document}